\documentclass[twoside,twocolumn,9pt]{article}
\usepackage{extsizes}
\usepackage[super,sort&compress,comma]{natbib}
\usepackage[version=3]{mhchem}
\usepackage[left=1.5cm, right=1.5cm, top=1.785cm, bottom=2.0cm]{geometry}
\usepackage{balance}
\usepackage{widetext}
\usepackage{times,mathptmx}
\usepackage{sectsty}
\usepackage{graphicx}
\usepackage{lastpage}
\usepackage[format=plain,justification=raggedright,singlelinecheck=false,font={stretch=1.125,small,sf},labelfont=bf,labelsep=space]{caption}
\usepackage{float}
\usepackage{fancyhdr}
\usepackage{fnpos}
\usepackage[english]{babel}
\usepackage{array}
\usepackage{droidsans}
\usepackage{charter}
\usepackage[T1]{fontenc}
\usepackage[usenames,dvipsnames]{xcolor}
\usepackage{setspace}
\usepackage[compact]{titlesec}
\usepackage{braket}	

\definecolor{cream}{RGB}{222,217,201}

\begin{document}

\pagestyle{fancy}
\thispagestyle{plain}
\fancypagestyle{plain}{

\renewcommand{\headrulewidth}{0pt}
}

\makeFNbottom
\makeatletter
\renewcommand\LARGE{\@setfontsize\LARGE{15pt}{17}}
\renewcommand\Large{\@setfontsize\Large{12pt}{14}}
\renewcommand\large{\@setfontsize\large{10pt}{12}}
\renewcommand\footnotesize{\@setfontsize\footnotesize{7pt}{10}}
\makeatother

\renewcommand{\thefootnote}{\fnsymbol{footnote}}
\renewcommand\footnoterule{\vspace*{1pt}%
\color{cream}\hrule width 3.5in height 0.4pt \color{black}\vspace*{5pt}}
\setcounter{secnumdepth}{5}

\makeatletter
\renewcommand\@biblabel[1]{#1}
\renewcommand\@makefntext[1]%
{\noindent\makebox[0pt][r]{\@thefnmark\,}#1}
\makeatother
\renewcommand{\figurename}{\small{Fig.}~}
\sectionfont{\sffamily\Large}
\subsectionfont{\normalsize}
\subsubsectionfont{\bf}
\setstretch{1.125} 
\setlength{\skip\footins}{0.8cm}
\setlength{\footnotesep}{0.25cm}
\setlength{\jot}{10pt}
\titlespacing*{\section}{0pt}{4pt}{4pt}
\titlespacing*{\subsection}{0pt}{15pt}{1pt}

\fancyfoot{}
\fancyfoot[LO,RE]{\vspace{-7.1pt}}
\fancyfoot[CO]{\vspace{-7.1pt}\hspace{13.2cm}}
\fancyfoot[CE]{\vspace{-7.2pt}\hspace{-14.2cm}}
\fancyfoot[RO]{\footnotesize{\sffamily{1--\pageref{LastPage} ~\textbar  \hspace{2pt}\thepage}}}
\fancyfoot[LE]{\footnotesize{\sffamily{\thepage~\textbar\hspace{3.45cm} 1--\pageref{LastPage}}}}
\fancyhead{}
\renewcommand{\headrulewidth}{0pt}
\renewcommand{\footrulewidth}{0pt}
\setlength{\arrayrulewidth}{1pt}
\setlength{\columnsep}{6.5mm}
\setlength\bibsep{1pt}

\makeatletter
\newlength{\figrulesep}
\setlength{\figrulesep}{0.5\textfloatsep}

\newcommand{\topfigrule}{\vspace*{-1pt}%
\noindent{\color{cream}\rule[-\figrulesep]{\columnwidth}{1.5pt}} }

\newcommand{\botfigrule}{\vspace*{-2pt}%
\noindent{\color{cream}\rule[\figrulesep]{\columnwidth}{1.5pt}} }

\newcommand{\dblfigrule}{\vspace*{-1pt}%
\noindent{\color{cream}\rule[-\figrulesep]{\textwidth}{1.5pt}} }

\makeatother

\twocolumn[
  \begin{@twocolumnfalse}
\vspace{3cm}
\sffamily
\begin{tabular}{m{4.5cm} p{13.5cm} }

 & \noindent\LARGE{\textbf{Phase-modulated electronic wave-packet interferometry reveals high resolution vibronic spectra of free Rb atoms and Rb*He molecules}} \\
\vspace{0.3cm} & \vspace{0.3cm} \\

 & \noindent\large{Lukas Bruder,$^{\ast}$\textit{$^{a}$} Marcel Mudrich,\textit{$^{a}$} and Frank Stienkemeier\textit{$^{a}$}} \\

 & \noindent\normalsize{Phase-modulated wave-packet interferometry is combined with mass-resolved photoion detection to investigate rubidium atoms attached to helium nanodroplets in a molecular beam experiment. The spectra of atomic Rb electronic states show a vastly enhanced sensitivity and spectral resolution when compared to conventional pump-probe wave-packet interferometry. Furthermore, the formation of Rb*He exciplex molecules is probed and for the first time a fully resolved vibrational spectrum for transitions between the lowest excited $5\Pi_{3/2}$  and the high-lying electronic states $2^2\Pi$, $4^2\Delta$, $6^2\Sigma$ is obtained and compared to theory. The feasibility of applying coherent multidimensional spectroscopy to dilute cold gas phase samples is demonstrated in these experiments.
} \\

\end{tabular}

 \end{@twocolumnfalse} \vspace{0.6cm}

  ]

\renewcommand*\rmdefault{bch}\normalfont\upshape
\rmfamily
\section*{}
\vspace{-1cm}

\footnotetext{\textit{$^{a}$~Physikalisches Institut, Universit\"at Freiburg, 79104 Freiburg, Germany. E-mail: lukas.bruder@physik.uni-freiburg.de\,.}}

\section{Introduction}
Wave-packet interferometry (WPI) is a fundamental and versatile tool to study and control the quantum dynamics in a wide range of target systems
\cite{fushitani_applications_2008, ohmori_wave-packet_2009, stienkemeier_coherence_1999, prakelt_phase_2004,  Mudrich:2008, mauritsson_attosecond_2010}. 
In these experiments, two wave-packets (WPs) prepared by identical but temporally separated optical pulses interfere constructively or destructively depending on their relative phase. 
The observation of such interferences is the prerequisite in coherent time-resolved spectroscopy or coherent control experiments\cite{fushitani_applications_2008, ohmori_wave-packet_2009}. 
The success of this technique usually relies on the precise control of the relative phase between pump and probe pulses and phase stabilization methods are required if a complete description of the coupled electronic and nuclear dynamics is desired
\cite{ohmori_wave-packet_2009}. 

Likewise, precise phase control is a key issue in higher order schemes such as optical multidimensional spectroscopy\cite{schlau-cohen_two-dimensional_2011}. 
Several phase stabilization methods have been developed 
\cite{de_boeij_phase-locked_1995, lepetit_linear_1995, tian_femtosecond_2003, brixner_phase-stabilized_2004, cowan_two-dimensional_2004, zhang_optical_2005, li_multiphoton_2007, tekavec_fluorescence-detected_2007, bristow_versatile_2009, nardin_multidimensional_2013, turner_invited_2011, rehault_two-dimensional_2014} 
to make two-dimensional electonic spectroscopy (2DES) accessible to the visible \cite{hybl_two-dimensional_1998}
 and ultraviolet optical range
\cite{tseng_two-dimensional_2009}. 
2DES has been applied to systems ranging from atomic vapors\cite{tian_femtosecond_2003, tekavec_fluorescence-detected_2007} to complex molecular systems in the liquid phase\cite{brixner_two-dimensional_2005, nemeth_tracing_2009} 
and bulk semiconductor nanostructures\cite{cundiff_optical_2012}. 
However, complementary investigations of dilute cold targets in molecular beams are yet missing. 

Recently, the Marcus group has established a collinear passively phase-stabilized 2DES scheme based on continuous acousto-optic phase modulation combined with lock-in demodulation (PM2D spectroscopy)\cite{tekavec_fluorescence-detected_2007}.
A variant of this technique has been implemented by the Cundiff group\cite{nardin_multidimensional_2013}.
PM2D spectroscopy has been combined with fluorescence and photocurrent detection\cite{tekavec_fluorescence-detected_2007, nardin_multidimensional_2013, karki_coherent_2014} and the applicability to highly dilute solutions was demonstrated\cite{lott_conformation_2011}.
The signal recovery capabilities of lock-in amplification and the incorporation of incoherent observables make this approach an ideal candidate for single-molecule or molecular beam studies. 
Initially, the phase modulation (PM) technique was demonstrated in a WPI experiment of atomic Rb vapor using fluorescence detection (PM-WPI)\cite{tekavec_wave_2006}. 
In the current work, we combine this PM-WPI scheme with mass-resolved photoion detection to investigate a dilute molecular sample in a supersonic beam setup. 
	
Helium nanodroplet isolation has been established as a unique technique for spectroscopic studies at millikelvin temperatures  \cite{Toennies:2004,Stienkemeier:2001,Choi:2006}. Atoms, molecules as well as complexes or clusters can be isolated in a weakly perturbing helium environment providing an ideal matrix where the spectroscopic resolution is significantly improved; not only in comparison with experiments in room temperature solvents\cite{Wewer:2004,Wewer:2005}, but also when compared to other cryogenic matrices or clusters \cite{Dvorak:2012}. 
Time-resolved experiments using femtosecond (fs) lasers\cite{Stienkemeier:2006} have focused on various photo dynamical processes of doped He droplets.
These include vibrational WP propagation \cite{Claas:2006,Claas:2007}, complex formation\cite{Droppelmann:2004, Giese:2012}, fragmentation dynamics\cite{Stienkemeier:2006}, desorption \cite{Vangerow:2014, Vangerow:2015} and various aspects of energy dissipation and decoherence \cite{Schlesinger:2010, Gruner:2011}.

Alkali metal atoms and small complexes, which are weakly bound at the surface of the droplets, were among the first species to be studied by electronic spectroscopy in He droplets\cite{stienkemeier:1995}. 
Upon excitation to the lowest excited P state, the enhanced interaction to helium leads to the formation of alkali-helium exciplex molecules which desorb off the droplet and can be probed in the gas phase\cite{Mudrich:2008, Droppelmann:2004, Giese:2012, Reho:1997, Schulz:2001}. 
The formation dynamics has been discussed for different alkali exciplexes (Na*He, K*He , Rb*He) \cite{Giese:2012, Reho:1997, Schulz:2001, Bruehl:2001, Fechner:2012}. 
Moreover, comperative studies using He droplets formed of the fermionic isotope $^3$He have been performed in order to probe the influence of superfluidity of the droplets\cite{Droppelmann:2004}. 

WPI experiments for the first time enabled to access the vibrational energies of such unstable complexes \cite{Mudrich:2008}. 
However, since only differences between individual vibrational levels are accessible using this technique, the assignment of vibrational states could not been done rigorously, and in particular the role of higher-lying electronic states correlating to the 5D atomic asymptote had not been included. 
Later, femtosecond pump - picosecond probe experiments have probed the dynamics of individual vibronic levels\cite{Giese:2012}. 
These measurements confirmed the previous rough assignment; however the spectral resolution stayed behind the one achieved using the WPI technique. 
In the current work, we present a new experimental approach to study the vibronic structure of cold isolated molecular samples by PM-WPI using photoion detection. 
The new PM technique both provides a direct access of vibronic energies, and a vastly enhanced spectral resolution which allows us to determine a large number of contributing states. In this way the involved potentials can be accurately tested on a level beyond that of current \emph{ab initio} calculations.

\section{He droplet beam generation and exciplex formation}
\begin{figure}[h]
\centering
  \includegraphics[width=0.9\columnwidth]{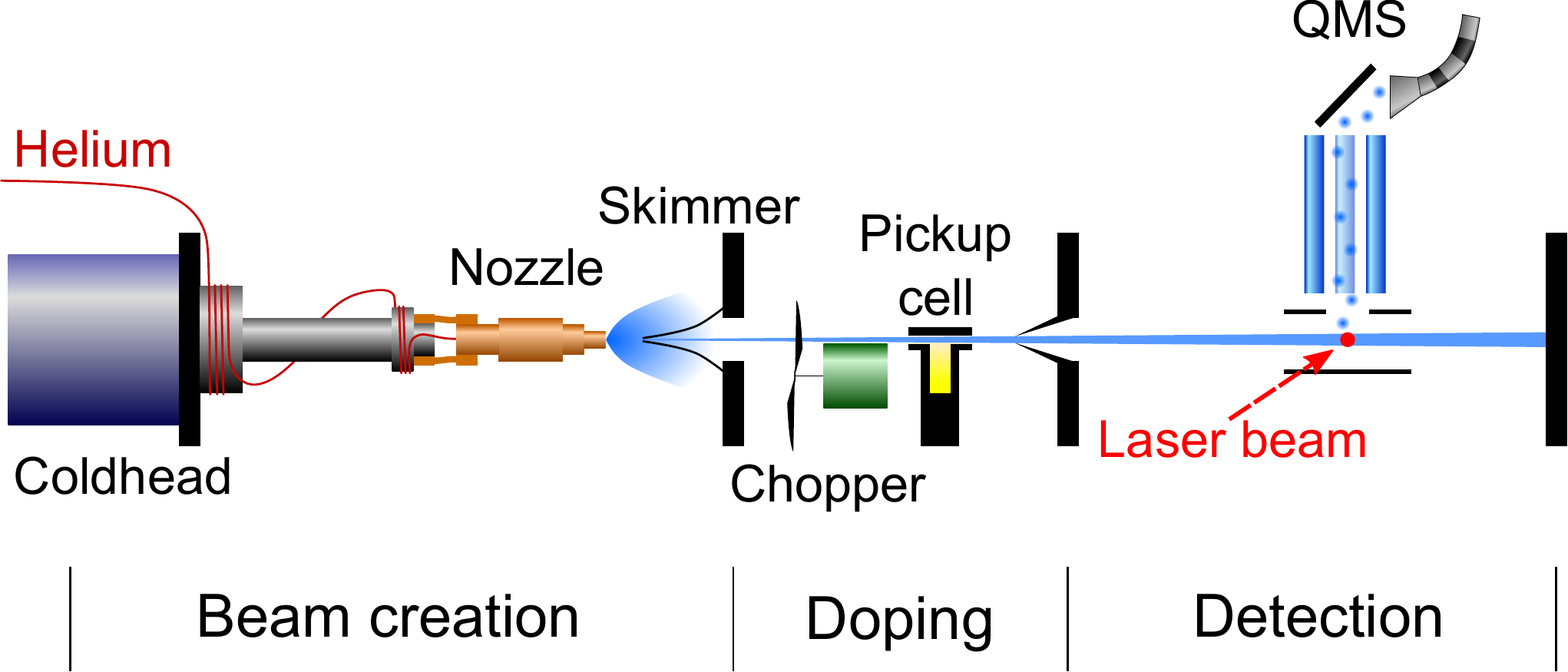}
  \caption{Schematic illustration of the molecular beam machine. A helium cluster beam traverses differentially pumped vacuum chambers housing the droplet source, pick-up cell and detector, respectively. }
  \label{fgr:dropletmachine}
\end{figure}
The molecular beam apparatus comprises of a differentially pumped vacuum system  where the formation of droplets, doping of droplets, and laser interaction and detection are individually housed and separated by gate valves (Fig.~\ref{fgr:dropletmachine}). The droplets are formed in a continuous flow  supersonic expansion from a nozzle 5\,$\mu m$ in diameter, attached to a closed cycle refrigerator. Expansion conditions are: nozzle temperature $T_0=17$\,K and backing pressure $p_{0}=50$\,bar , which corresponds to a mean cluster size of He$_{N}$, $N=20,000$\cite{Stienkemeier:2006}. Doping of the He droplets with Rb atoms is achieved in a heated vapor cell which is stabilized to a temperature $T_{Rb}=358K$. 
At this temperature, the Rb vapor pressure is such that on average only one Rb atom is attached to each droplet. The fs laser pulses intersect the doped droplet beam perpendicularly inside the ion extraction region of a commercial quadrupole mass spectrometer (QMS). 
Since atoms and exciplexes tend to desorb off the droplets upon electronic excitation\cite{Vangerow:2015}, the photo-ionized exciplexes are detected on their isotope molecular masses, well separated from that of bare Rb atoms.  

\begin{figure}[h]
\centering
  \includegraphics[width=0.99\columnwidth]{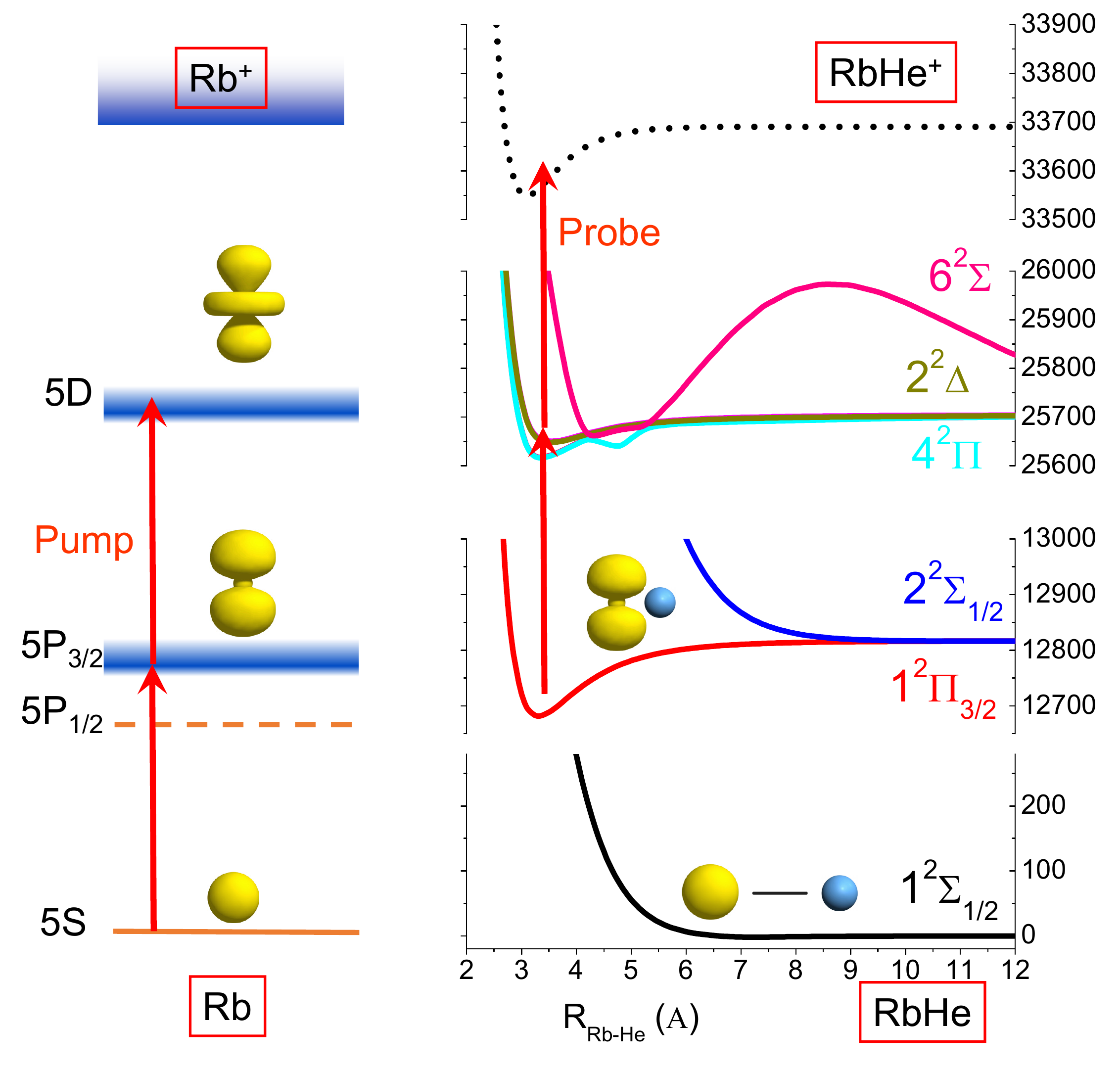}
  \caption{Pump-Probe excitation scheme. The Rb $5P_{3/2}$ and $5D$ states are coherently excited. The broadening of states induced by the helium environment is indicated. The relevant Rb*He pair potentials of the formed exciplexes and the probe process are illustrated in the right panel.}
  \label{fgr:levelscheme}
\end{figure}

The formation of Rb*He molecules is based on the attractive interaction of Rb excited states with He atoms, mainly due to a missing Pauli repulsion in the nodal region of extended electron orbitals having angular momentum $l>0$. 
However, a complete understanding of the formation and desorption process of metal-He exciplexes at He droplets is still missing. 
Different models have been discussed in order to interpret the formation of alkali helium exciplexes and the measured time dependent signals. 
Tunneling processes have been modeled \cite{Reho:2000,Reho2:2000,Loginov:2007,loginov_excitation_2015} and direct laser-excitation of bound states in the excited pair potential has been discussed \cite{Pascale:1983,Fechner:2012,Vangerow:2014,Loginov:2014}. 
In Fig.~\ref{fgr:levelscheme} the potentials of the relevant states for our experiment are plotted and the pump - probe scheme is illustrated by vertical arrows.
   
\section{Phase-modulated wave-packet interferometry}
Conventional pump-probe spectroscopy allows to follow the population dynamics in a quantum system at real-time, however not the dynamics of coherences or coupled electronic-vibrational dynamics. 
(Electronic) WPI experiments, on the contrary, give access to the full information content by employing a coherent excitation scheme involving phase-related pump-probe pulses\cite{ohmori_wave-packet_2009}. 
In this approach, two WPs separately excited by pump and probe pulses interfere with each other, giving rise to constructive and destructive interference effects in the underlying quantum system. 
In the regime of weak optical perturbation, WPI is equivalent to quantum beat measurements, Ramsey fringes in the time domain\cite{noordam_ramsey_1992, christian_rubidium_1993}, Fourier spectroscopy using fs pulses\cite{bellini_two-photon_1997}, or temporal coherent control\cite{blanchet_temporal_1997}.  

\begin{figure}[h]
\centering
  \includegraphics[width=0.99\columnwidth]{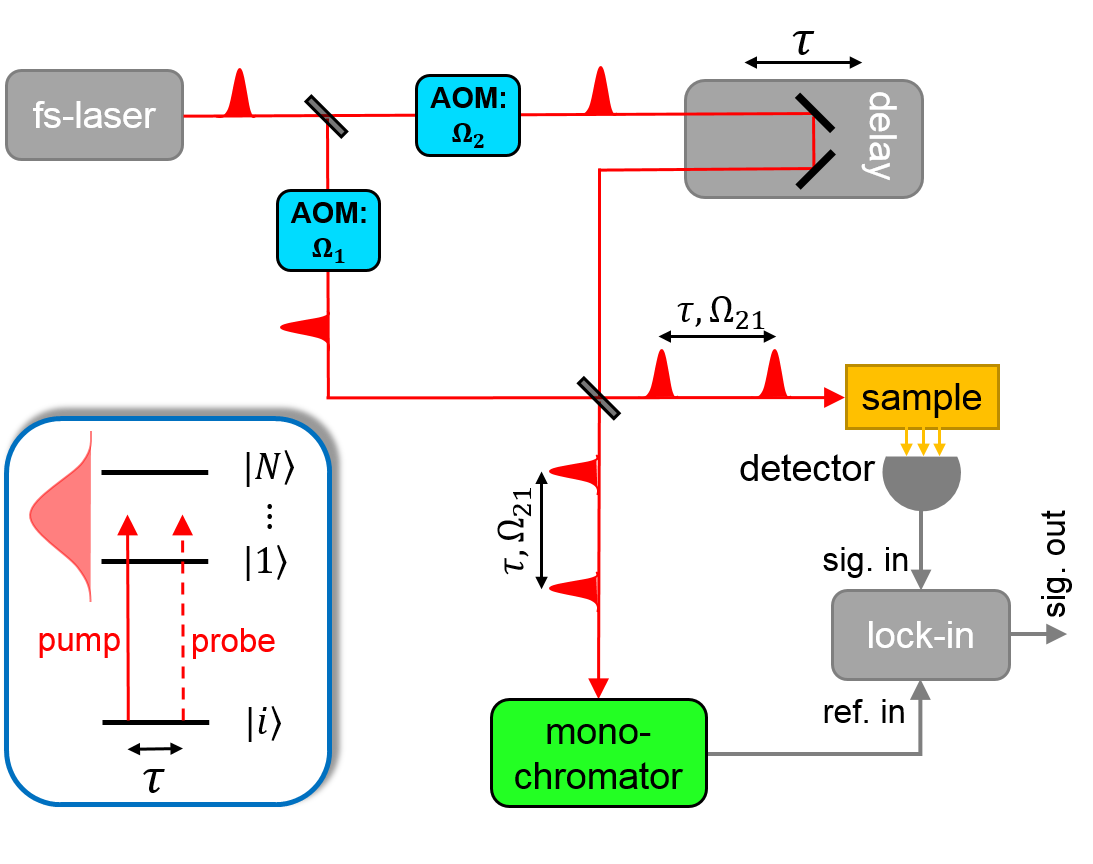}
  \caption{Simplified diagram of the PM-WPI scheme. Fs pump-probe pulse sequences are generated in a Mach-Zehnder interferometer. AOMs impart a continuous phase-modulation onto the optical pulses. A monochromator is used to construct a reference signal for lock-in demodulation of the acquired WPI data. The box labeled \textit{sample} represents the molecular beam machine. The inset shows the generalized WPI excitation scheme.}
  \label{fgr:PM_scheme}
\end{figure}

In more detail, we consider a simple system consisting of a few discrete energy eigenstates $\ket{i},\ket{1},...,\ket{N}$ as shown in the inset of Fig.\,\ref{fgr:PM_scheme}. 
The pump pulse excites a coherent superposition of the initial state $\ket{i}$ and excited states $\ket{n}$ lying within the laser bandwidth. 
This WP evolves freely in time until the probe pulse excites a second WP. 
The interference of the two WPs induces a modulation of the excited state populations with respect to the pump-probe delay $\tau$. 
Probing the excited state populations while systematically scanning $\tau$ yields a typical quantum beat signal\cite{scherer_fluorescence-detected_1991, tekavec_wave_2006}
\begin{equation} \label{Eq:BeatSignal}
	S(\tau) = const. + \sum_{n=1}^N A_n \cos(\omega_{ni} \tau),
\end{equation}
where $\omega_{ni}=(E_n-E_i)/\hbar$ corresponds to the energy differences of the optical transitions from level $\ket{i}$ into $\ket{n}$. 
$A_n \propto |D_{ni}|^2\tilde{E}^2(\omega_{ni})$ represents the transition probability for each $\ket{n} \leftarrow \ket{i}$ transition, 
where $\tilde{E}(\omega)$ is the Fourier transform of the single pulse electric field (assuming identical pump and probe pulses) and $D_{ni}$ represents the transition dipole moment of the relevant degrees of freedom (electronic, vibrational and rotational). 
For simplicity, temporal overlap between pump and probe pulses is neglected and their relative carrier-envelope phase is set to zero. 
The constant offset in Eq.\,\ref{Eq:BeatSignal} originates from excitation pathways involving interactions with pump or probe pulse only and thus does not depend on the pump-probe delay. 

The PM-WPI approach of the Marcus group effectively extracts the pure two-pulse interference contribution of Eq.\,\ref{Eq:BeatSignal} by combining acouto-optic PM with lock-in demodulation. 
A detailed description of the PM-WPI scheme is given by Tekavec et al\cite{tekavec_wave_2006}. 
We apply this technique in an almost identical optical setup (schematically shown in Fig.\,\ref{fgr:PM_scheme}). 
Briefly, pump-probe pulse pairs of adjustable delay are generated in a Mach-Zehnder type interferometer  and are collinearly superimposed afterwards. 
An acousto-optic modulator (AOM) placed in each branch of the interferometer shifts the carrier frequency of pump and probe pulses individually by a small distinct amount. 
Phase-locked driving of the two AOMs thus imparts a well-defined difference frequency (denoted $\Omega_{21}$) onto the two-pulse intensity function of the copropagating pulses. 
In principle, this is equivalent to shot-to-shot phase-cycling, where the relative phase between pump and probe pulses is incremented by $\Delta\phi=\Omega_{21}T_{rep}$ ($1/T_{rep}$ being the laser repetition rate) between each laser shot\cite{nardin_multidimensional_2013}. 
Eventually, this results in excited state populations in the sample that exhibit, besides the typical quantum beat of Eq.\,\ref{Eq:BeatSignal}, an additional continuous modulation with the difference frequency of the AOMs 
\begin{equation}	\label{Eq:ModBeatSignal}
	S_{in}(\tau,t) = const. + \sum_{n=1}^N A_n \cos(\omega_{ni} \tau - \Omega_{21} t).
\end{equation} 
Here the assumption has been made that $\Omega_{21} \ll 1/T_{rep}$. 
Thus $t$ represents a quasi-continuous time variable. 
The two-pulse intensity function is spectrally filtered by a monochromator in a separate beam, yielding a reference signal for phase-synchronous detection, 
\begin{equation} \label{Eq:RefSignal}
	R(\tau,t) = R_0 \cos(\omega_{M} \tau - \Omega_{21} t).
\end{equation}
Here the band-pass filter function of the monochromator is centered at the frequency denoted $\omega_M$. 
Applying the signal of Eq.\,\ref{Eq:ModBeatSignal} to a lock-in amplifier while referencing the device with the signal of Eq.\,\ref{Eq:RefSignal} results in an output signal of the form 
\begin{equation} \label{Eq:DMBeatSignal}
	S_{out}(\tau) = \sum_{n=1}^N A_n \cos\left[ \left(\omega_{ni}-\omega_M\right) \tau \right].
\end{equation}

It is straight forward to apply these considerations to target systems of different energetic structure. 
In the current work, we investigate atomic Rb as well as Rb*He exciplex molecules by means of photoionization and subsequent mass-resolved ion detection. 
In the atomic case, we follow excitation pathways where coherent superpositions of the 5S$_{1/2}$ and 5P$_{3/2}$ states as well as of the 5P$_{3/2}$ and 5D$_{5/2,\,3/2}$ states are induced, as indicated in Fig.\,\ref{fgr:levelscheme}. 
In the molecular case, we prepare coherent superpositions between vibrational states of the potential energy curves (PECs) correlating to the 5P$_{3/2}$ and 5D$_{5/2,\,3/2}$ asymptotic atomic states, also indicated in Fig.\,\ref{fgr:levelscheme}. 
In both cases, the phase-related pump-probe sequence eventually prepares a stationary population state in the target system which is ionized by subsequent absorption of additional photons within the same probe pulse or by consecutive pulses. 
In analogy to fluorescence detection\cite{tekavec_wave_2006}, the ion yield reflects the mean population probability of the excited states and is thus directly proportional to the individual beat components of Eq.\,\ref{Eq:DMBeatSignal}. 

When comparing the outcome of a conventional WPI experiment (Eq.\,\ref{Eq:BeatSignal}) with the outcome of a PM-WPI experiment (Eq.\,\ref{Eq:DMBeatSignal}), several advantages of the latter method become obvious. 
In PM-WPI experiments, one measures the actual optical transition frequencies relative to the monochromator frequency, resulting in a strongly undersampled quantum beat signal. 
Therefore, the same information content can be inferred from a much smaller amount of sampling points. 
Moreover time-delay jitter due to mechanical instabilities within the interferometer scale to a significantly less extent and laboratory noise in general is strongly suppressed by the lock-in demodulation process. 
As mentioned above, the phase modulation scheme introduces shot-to-shot or \textit{dynamic} phase-cycling\cite{nardin_multidimensional_2013}. 
This isolates excitation pathways that result in an $\Omega_{21}$-modulation of the final population state. 
For instance pathways ending in a population state involving interactions with only one single pulse or multiple but not phase-related pulses do not contribute to the signal of Eq.\,\ref{Eq:DMBeatSignal}.

For the PM-WPI optical setup we employ a titanium:sapphire oscillator producing pulses of about 200\,fs duration and 70\,cm$^{-1}$ spectral width (FWHM) at a repetition rate of 80\,MHz. 
Average laser power is about 2\,W at the laser exit window, leading to pulse energies of 1.7\,nJ (pump) and 5.2\,nJ (probe) before entering the vacuum apparatus. 
The laser beam is focused with a lens of 150\,mm focal length, resulting in a beam diameter of 40\,$\mu$m at the interaction volume. 
For the presented results, we acquire time-domain interferograms by sampling the pump-probe delay in 30\,fs steps from 0 to 50\,ps in case of the atomic Rb measurements and in 50\,fs steps from -100 to 120\,ps for the Rb*He exciplex measurements. 
For each delay value, 1000 sample points are averaged in the atomic Rb measurements and 1500 sample points in the Rb*He exciplex measurements, respectively. 
The QMS signal is amplified before feeding it into the lock-in amplifier.   
The monochromator signal is acquired with an avalanche photo diode, band-pass filtered and amplified. 
Signal demodulation is done with a lock-in time constant of 30\,ms and 18\,dB roll-off. 
Demodulated in-phase (X) and in-quadrature (Y) signals are acquired simultaneously to reconstruct the complex-valued amplitude $Z=X+iY$. 
A discrete Fourier transform of $Z$  is performed after applying a Gaussian window function and zero-padding. 
The frequency axis of the obtained spectrum is shifted by the amount of the monochromator frequency in order to reconstruct the absolute transition frequencies. 
In case of the exciplex experiments, the frequency spectrum is additionally deconvoluted with the pump-probe autocorrelation function, analog to the procedure applied in Ref. \cite{scherer_fluorescence-detected_1991}. 
\begin{figure}[h]
\centering
  \includegraphics[width=0.9\columnwidth]{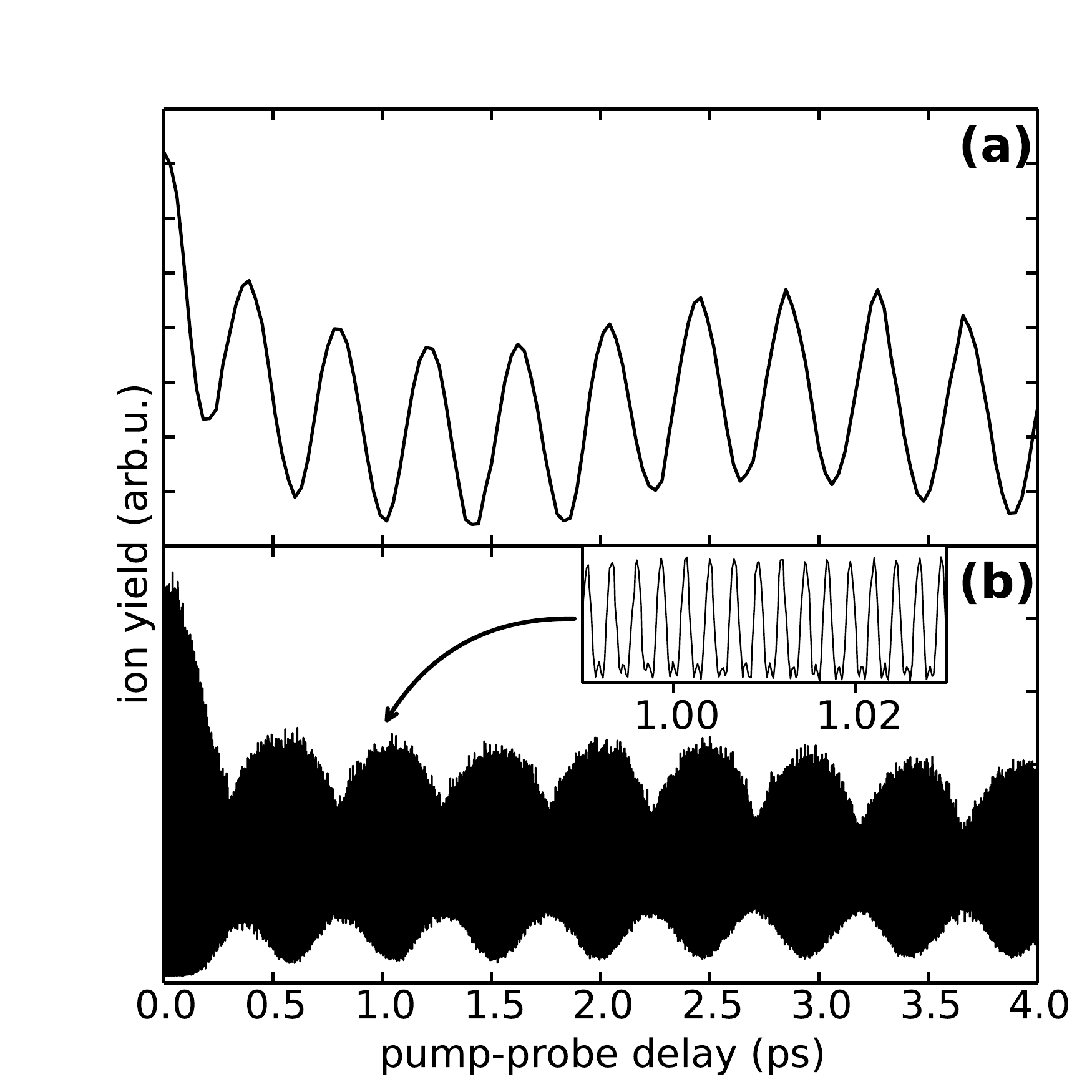}
  \caption{WPI measurements of Rb atoms attached to He droplets. (a) shows the Rb ion yield as a function of pump-probe delay obtained from a PM-WPI measurement (in-phase component) and (b) respective data obtained in a conventional WPI experiment. Comparing the signal frequencies in (a) with the inset of (b) highlights the drastic frequency downshifting effect inherent in the PM-WPI technique.}
  \label{fgr:Rb_interferogram}
\end{figure}
\begin{figure}[h]
\centering
  \includegraphics[width=0.8\columnwidth]{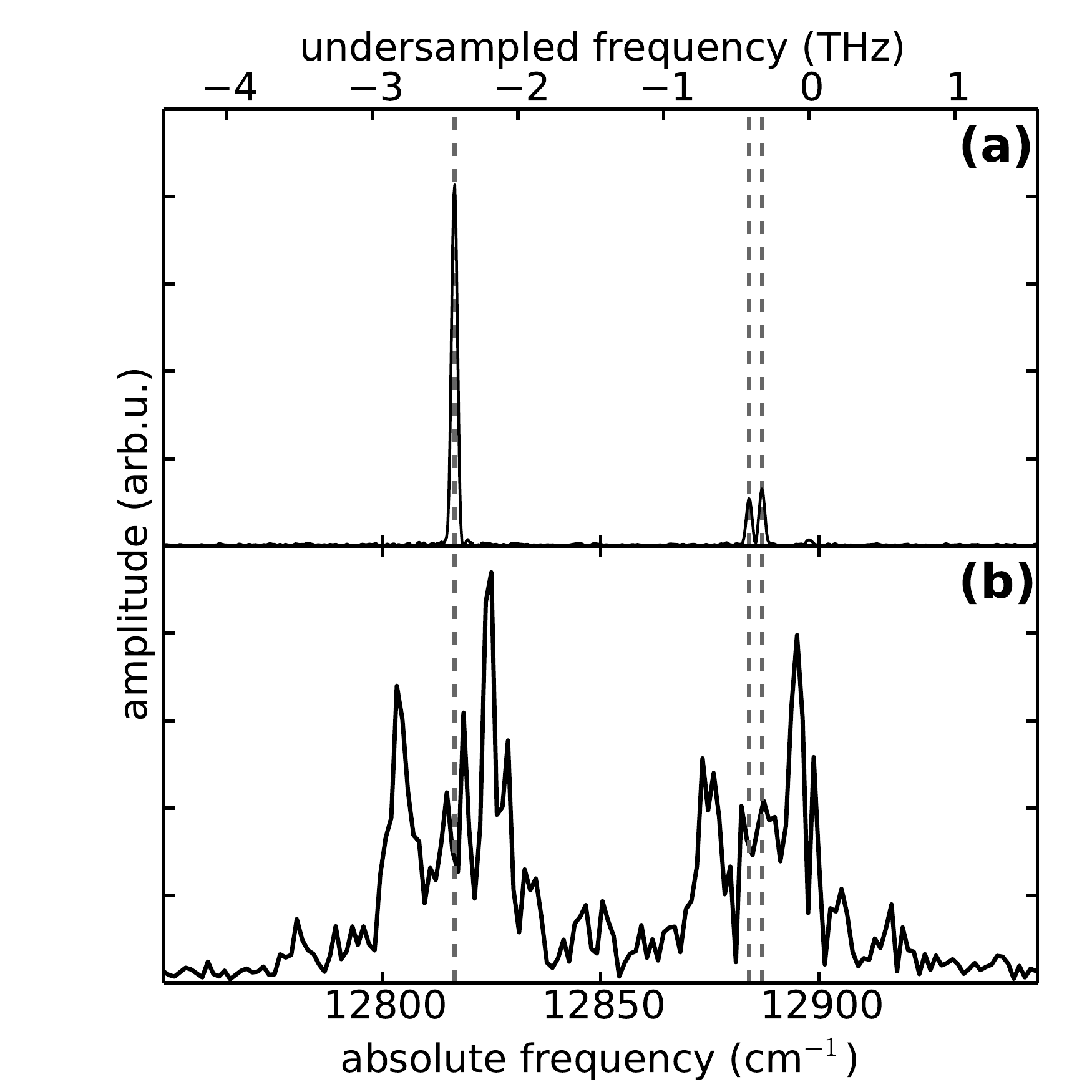}
  \caption{Discrete Fourier transform of the PM-WPI time domain data (a) and conventional WPI data (b), respectively, for measurements of Rb atoms attached to He droplets. The absolute frequencies given by the bottom scale are common for both graphs. The scale at the top shows the downshifted frequencies at which the pump-probe transient is detected in the PM scheme before being shifted by the monochromator frequency. Vertical dashed lines indicate the atomic Rb transition frequencies.}
  \label{fgr:Rb_DFT}
\end{figure}

\section{PM-WPI of atomic Rb attached to He droplets}
To demonstrate the advantages of the PM method, we show at first measurements of atomic Rb attached to He droplets which serves as a simple model system with only four electronic states involved, as illustrated in Fig.\,\ref{fgr:levelscheme}. 
For this purpose the mass spectrometer is tuned to the mass of the most abundant isotope $^{85}$Rb and the laser wavelength is set to 775\,nm. 
The 5S$_{1/2} \rightarrow $ 5P$_{3/2}$  and 5P$_{3/2}\rightarrow $ 5D$_{5/2,\,3/2}$ transitions are separated by 70.40\,cm$^{-1}$ and 67.44\,cm$^{-1}$, respectively, which fall into the bandwidth of the fs pulses. 
In the applied WPI scheme, interference among pump and probe excitation thus results in excited state populations that exhibit delay-dependent oscillations corresponding to the mentioned optical transition frequencies. 
We compare the PM-WPI data with conventional WPI measurements previously conducted in our group with the same equipment, under similar experimental conditions\cite{heister_unpublished_2008}. 
Fig.\,\ref{fgr:Rb_interferogram} shows the ion yield for the first 4\,ps of pump-probe delay for the PM-WPI (a) and conventional WPI case (b), respectively. 
Both interferograms exhibit an exaggeration of the ion yield around zero delay resulting from temporal overlap of pump and probe pulses. 
Fig.\,\ref{fgr:Rb_DFT} shows the discrete Fourier transforms of the time domain data. 
The pronounced peaks in (a) correspond precisely to the expected 5S$_{1/2} \rightarrow $ 5P$_{3/2}$  and 5P$_{3/2}\rightarrow $ 5D$_{5/2,\,3/2}$ transitions (shown as gray dashed lines). 
Here even the spin-orbit splitting of the 5D state ($\Delta$=2.96cm$^{-1}$) is resolved, whereas in (b) transition frequencies can be identified only qualitatively and the splitting of the 5D state is not resolved. 

Two effects are responsible for the drastic difference in signal quality achieved by the two experimental techniques. 
On the one hand, the PM-WPI data is measured in the rotating frame defined by the monochromator frequency which results in strong undersampling of the quantum beat frequencies. 
In the conventional WPI experiment (Fig.\,\ref{fgr:Rb_interferogram}\,(b)) the optical transition frequencies are fully sampled as shown in the inset, whereas in Fig.\,\ref{fgr:Rb_interferogram}\,(a) the pump-probe transients are measured in the rotating frame and thus downshifted from $\approx$380\,THz to below 3\,THz (refer to the top frequency scale in Fig.\,\ref{fgr:Rb_DFT}). 
This significantly reduces the demands on phase stability in the optical setup. 
In the presented conventional WPI experiment, phase errors have been partially compensated by tracking the interferogram of a HeNe laser while scanning the pump-probe delay whereas no phase corrections have been applied in the PM scheme. 
Nonetheless, the latter case exhibits a drastic improvement in resolution and an excellent signal-to-noise ratio. 
At the same time, less data points are required to obtain the same information: 
the PM-WPI data set (0-50\,ps pump-probe delay, 50\,fs step size) comprises only 1.6\,\% of the sampling points acquired in the fully sampled case (0-63\,ps pump-probe delay, 0.6\,fs step size). 

On the other hand, imprinting a well-defined modulation onto the WPI signal, allows utilizing efficient lock-in detection to extract weak signals from very large background contributions. 
This advantage of the PM method is particularly important when investigating highly dilute samples. 
In He droplet beam experiments, target densities are estimated to fall below 10$^8$\,cm$^{-3}$\,\cite{Stienkemeier:2006} which usually puts a severe constraint to the achievable signal-to-noise ratio; the lock-in amplification in the PM scheme, however, provides great background suppression.  

\section{PM-WPI of Rb*He exciplexes}
\subsection{Experimental results}
\begin{figure}[h]
\centering
  \includegraphics[width=\columnwidth]{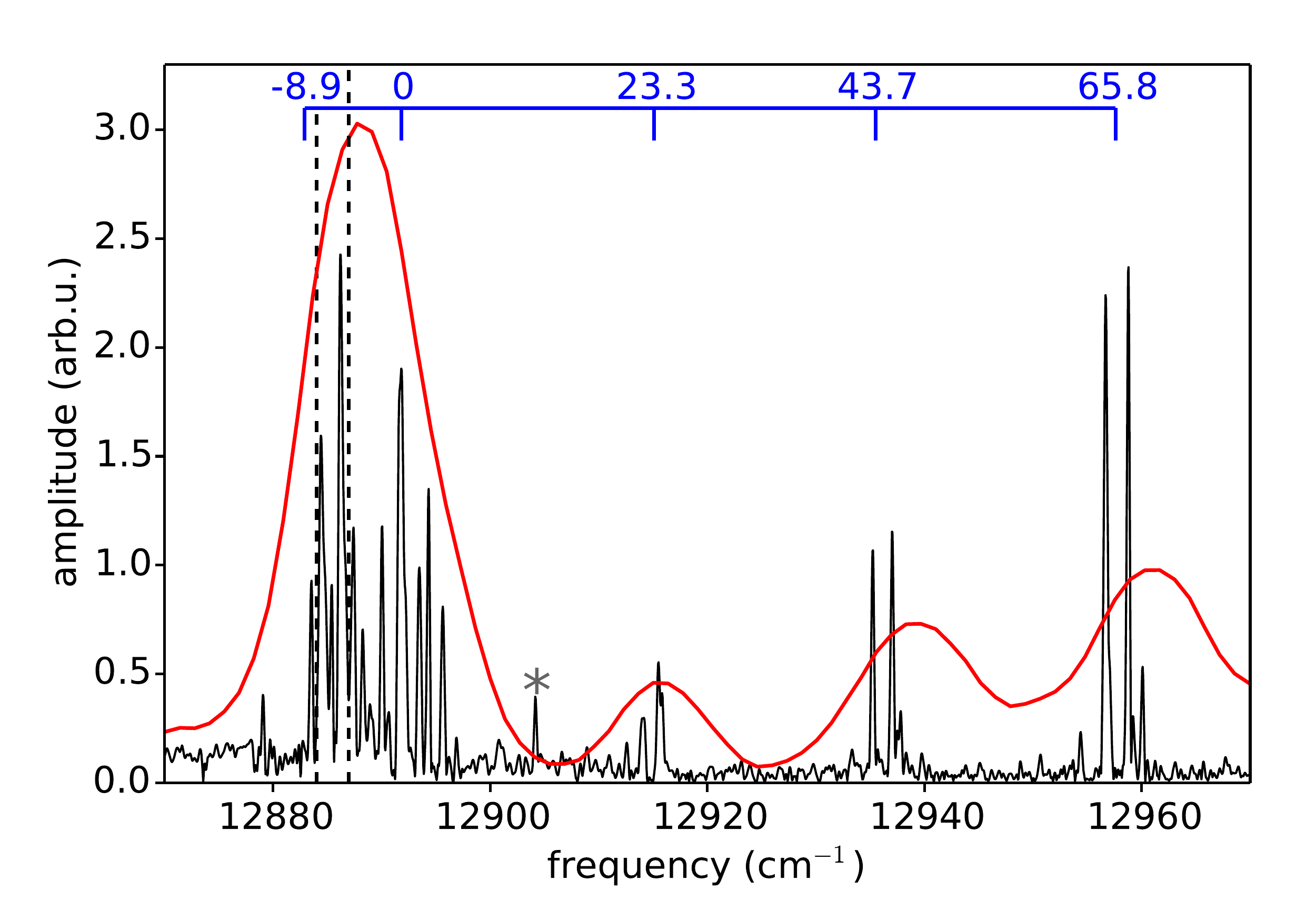}
  \caption{Rb*He exciplex spectrum obtained from PM-WPI (black) and fs pump ps probe (red) measurements\cite{Giese:2012}. Relative transition frequencies inferred from conventional WPI\cite{Mudrich:2008} are also shown (blue scale on top). Black dashed lines indicate the 5P$_{3/2}\rightarrow $ 5D$_{5/2,\,3/2}$ atomic Rb transitions. In the PM-WPI spectrum, one peak originating from laboratory noise appears at the employed monochromator frequency (asterisk).}
  \label{fgr:exciplex_comparison}
\end{figure}

After having established the accuracy and sensitivity of the PM-WPI method, we turn towards studying Rb*He exciplexes, the spectra of which have not yet been recorded with high resolution. 
PM-WPI measurements are conducted by tuning the mass spectrometer to 89\,amu and the laser wavelength to 774\,nm. 
Signal strengths are a factor of 20 smaller than in the atomic Rb case. 
In the PM-WPI scheme, vibronic transitions between the $1^2\Pi_{3/2}$ potential energy curve (PEC) correlating to the 5P$_{3/2}$ and the $2^2\Pi$, $4^2\Delta$, $6^2\Sigma$ PECs correlating to the 5D$_{5/2,\,3/2}$ atomic Rb states are probed (cf$.$ Fig.\,\ref{fgr:levelscheme}). 
Pure vibrational transitions within the same PEC or transitions among the $2^2\Pi$, $4^2\Delta$, $6^2\Sigma$ PECs are not detected due to the phase-cycling condition inherent in the PM scheme. 
A Fourier transform of the pump-probe transient yields a fully resolved vibrational spectrum of the exciplex molecule (Fig.\,\ref{fgr:exciplex_comparison}) with a spectral resolution of 0.3\,cm$^{-1}$ (FWHM). 
Comparing the spectral lines with the position of the 5P$_{3/2}\rightarrow $ 5D$_{5/2,\,3/2}$ atomic Rb transitions (black dashed lines), an intuitive interpretation is possible: 
the closely spaced spectral lines in the vicinity of the atomic transitions involve mainly vibrational modes energetically close to the dissociation limits. 
Lines at larger frequencies correspond to transitions of lower lying vibrational modes in the $1^2\Pi_{3/2}$ electronic state and the line splitting originates from the congestion of the higher-lying PECs. 

Let us compare our data with previous photoionization studies of the Rb*He exciplex. 
In a conventional WPI experiment\cite{Mudrich:2008} the signal quality was not sufficient to deduce the absolute transition frequencies through a direct Fourier analysis; relative distances between vibrational levels were deduced instead. 
To compare with our PM-WPI results, we assign the relative frequencies obtained in Ref. \cite{Mudrich:2008} to the distances between spectral features in our study (Fig.\,\ref{fgr:exciplex_comparison}, blue scale on top). 
We find good agreement albeit our spectrum exhibits a factor 10 better resolution and allows a more unambiguous interpretation. 
A different approach for achieving vibrationally resolved data on Rb*He exciplex molecules has been pursued in a femtosecond pump - picosecond probe experiment\cite{Giese:2012}. This study allowed to deduce absolute transition frequencies although at the cost of resolution which is factor 30 lower than achieved in our measurement. 
After applying a red shift of 2\,cm$^{-1}$ to the corresponding data (red spectrum in Fig.\,\ref{fgr:exciplex_comparison}) good agreement with our spectrum is obtained despite the different measurement schemes. 
The systematic shift can be attributed to coarse calibration of the pulse shaper that was used to generate the ps probe pulses. 
Calibration errors in the PM-WPI data can be excluded since calibration measurements with a Rb vapor cell have been carried out regularly during the measurements. 

\begin{figure}[h]
\centering
  \includegraphics[width=\columnwidth]{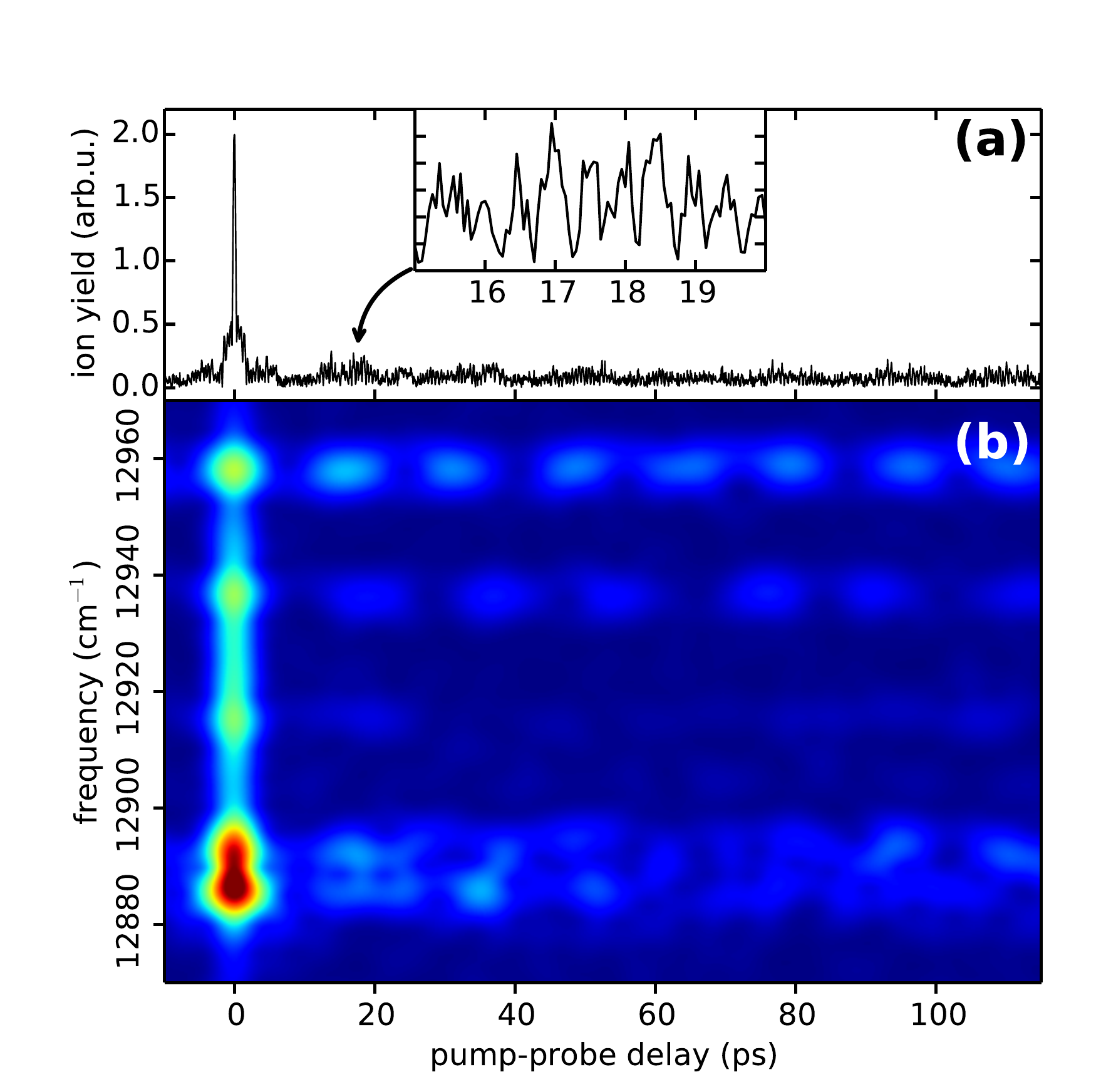}
  \caption{Time-domain data of the Rb*He exciplex (a) and spectrogram representation obtained by short-time Fourier analysis (b). In the spectrogram, the time evolution of the most prominent spectral features is directly visible.}
  \label{fgr:spectrogram}
\end{figure}
Previously, two types of dynamics have been observed when forming Rb*He exciplexes on the surface of He droplets. 
On the one hand, a fast process on the few ps time scale is observed\cite{Mudrich:2008} which can be attributed to an impulsive response of the He droplet environment to the electronic excitation of the Rb atom. 
On the other hand, slow relaxation of vibrational modes on the ns time scale is observed and attributed to slow vibrational dephasing induced by the droplet environment\cite{Giese:2012}. 
Fig.\,\ref{fgr:spectrogram} shows the time-domain data of the Rb*He exciplex obtained in our experiment. 
The absolute value of the complex-valued pump-probe transient is shown in (a); recorded (undersampled) quantum beats are  enlarged in the inset. 
The pronounced exaggeration of the ion yield around zero pump-probe delay (narrow spike in Fig.\,\ref{fgr:spectrogram}\,(a)) is due to the temporal overlap of pump and probe pulses. 
Outside of this region, a fast decay of the ion yield within 1.5\,ps is visible (shoulder in Fig.\,\ref{fgr:spectrogram}\,(a)) which is in accordance with the observed dynamics in Ref.\,\cite{Mudrich:2008}. 
In order to make the time evolution of the most prominent spectral features directly accessible, a spectrogram representation of the time domain data obtained by short-time Fourier analysis (Gaussian window function of 4\,ps FWHM, 0.5\,ps step size) is shown in Fig.\,\ref{fgr:spectrogram}\,(b).  
The short-time Fourier transform results, however, in a temporal blurring. 
Thus the discussed fast decay is not observable in Fig.\,\ref{fgr:spectrogram}\,(b) but several beat features originating from closely spaced spectral lines are well resolved. 
Traces along these beats indicate a slow relaxation presumably on the ns time scale (not shown) which is in accordance with the results of Ref.\cite{Giese:2012}. 
However, a longer time window is needed to provide a reliable interpretation in our data. 

\subsection{Comparison with theory}
\begin{figure}[h]
\centering
  \includegraphics[width=0.97\columnwidth]{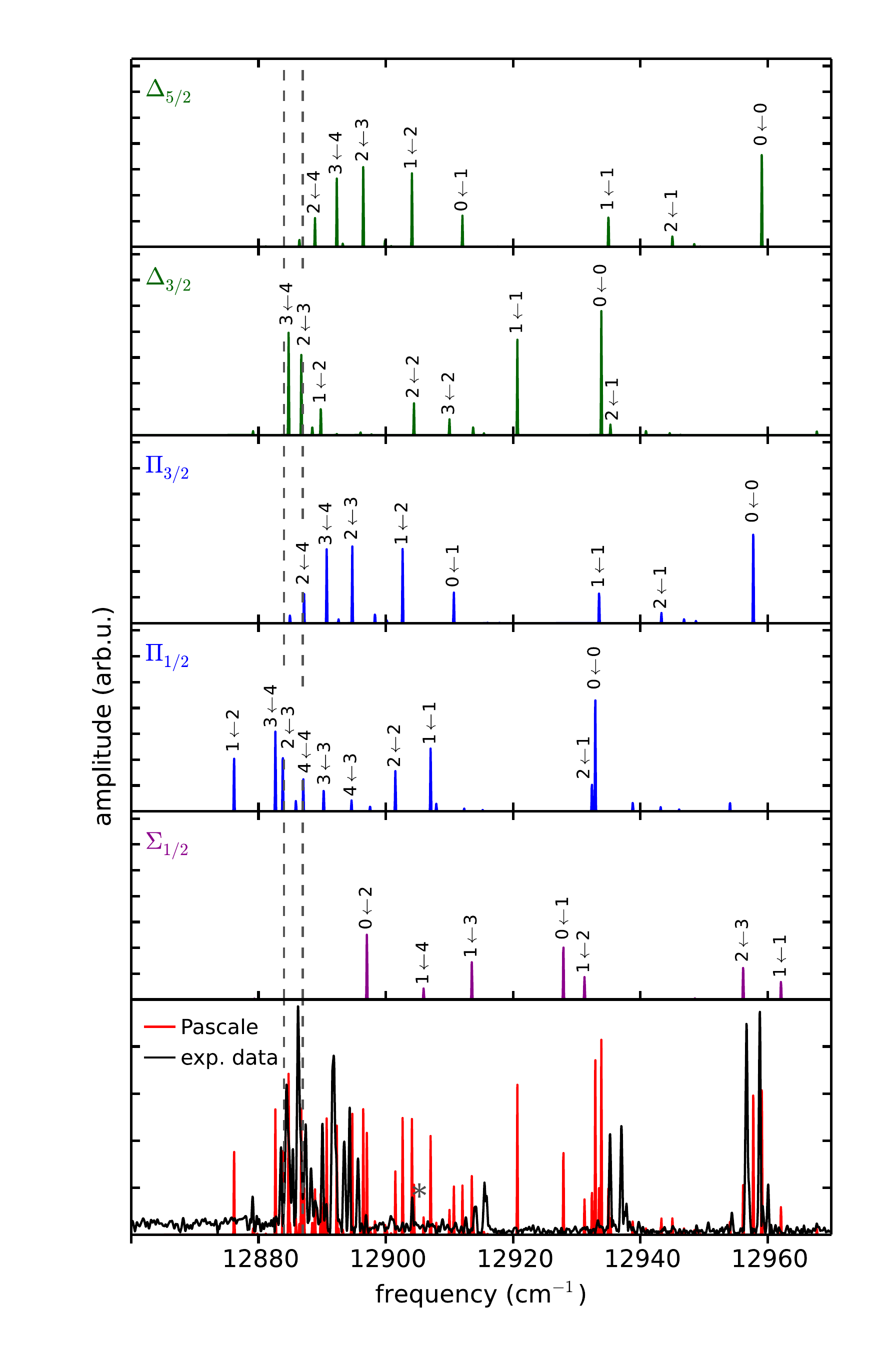}
  \caption{Experimental Rb*He exciplex spectrum compared with calculations based on the potentials of Pascale\cite{Pascale:1983}. The five upper panels show the calculated FCFs for transitions from the $1^2\Pi_{3/2}$ lower potential separately to each higher-lying potential (label). Adding up all five spectra yields the red spectrum in the bottom panel which is compared to the experimental data (black). Transitions have been convoluted with a Gaussian of 0.3\,cm$^{-^1}$ and blue shifted by 4\,cm$^{-1}$. For simplicity only most prominent peaks are labeled.}
  \label{fgr:exciplex_Pascale}
\end{figure}
Despite the availability of more recent \textit{ab initio} PECs of the Rb*He exciplex\cite{blank_m_2012, chattopadhyay_spectroscopic_2012, bouhadjar_rubidium_2014}, we compare our data with the semi-empirical potentials calculated by Pascale\cite{Pascale:1983}, which is currently the only model comprising highly excited electronic states. 
Pascale's pair potentials have proven to accurately describe the energetic structure of alkali-helium exciplexes in many cases\cite{Reho:1997, Bruehl:2001, Giese:2012, loginov_excitation_2015}. 
In particular, results from recent high resolution zero electron kinetic energy (ZEKE) studies of the NaHe exciplex were in much better accordance with the Pascale PECs than with state-of-the-art \textit{ab initio} calculations\cite{loginov_excitation_2015}. 
To account for spin-orbit couplings, we apply the procedure described in Refs.\cite{Bruehl:2001, Reho2:2000} to the Pascale PECs. 
To calculate the WPI spectrum, Frank-Condon factors (FCFs) for transitions between the $1^2\Pi_{3/2}$ lower potential and the $2^2\Pi$, $4^2\Delta$, $6^2\Sigma$ upper PECs are evaluated employing the LEVEL 8.0 code of LeRoy\cite{leroy_level_2007}. 
Transition dipole moments are assumed to be constant. 
The obtained stick spectra are convoluted with a Gaussian of 0.3\,cm$^{-1}$ FWHM to account for the experimental resolution. 
Furthermore, a systematic blue shift of 4\,cm$^{-1}$ is applied, which is a reasonable adjustment for such type of numeric models. 
The resulting spectrum is shown in Fig.\,\ref{fgr:exciplex_Pascale} together with the experimental data. 
We find semi-quantitative agreement between experiment and theory, which is quite astonishing considering the degree of detail revealed in the experimental data and the high experimental resolution which certainly exceeds the precision of the Pascale PECs. 
For an assignment of spectral lines to specific vibronic transitions, we show separately in five sub panels the calculated transitions from the $1^2\Pi_{3/2}$ intermediate PEC to the individual upper potentials (labeled sub panels in Fig.\,\ref{fgr:exciplex_Pascale}). 
This allows a more precise interpretation of the exciplex spectrum than was possible in previous studies\cite{Mudrich:2008, Giese:2012}. 
Obviously, the two clusters of lines at higher frequencies correspond primarily to 0$\leftarrow$0 transitions and do not reflect the vibrational ladder of the $1^2\Pi_{3/2}$ potential, as has been thought before. 
This observation emphasizes the importance of including accurate higher-lying electronic states when performing WPI on this system or when probing via resonance enhanced multi photon ionization (REMPI). 

The exciplex spectrum has also been calculated including rotational transitions which yielded much poorer agreement with the experiment (not shown). 
Thus, rotational features seem to be missing in our measurement despite the fact that our spectral resolution is sufficiently high to reveal those. 
We assume that rotations are not observed because of efficient cooling by the He droplet environment. 
This is in accordance with previous vibronic WPI measurements of Rb dimers formed on the surface of He droplets which also lacked spectroscopic signatures of rotational transitions\cite{Gruner:2011}.

\section{Conclusion}
With this work we demonstrate for the first time the applicability of PM-WPI to a molecular sample in the gas phase at very low particle densities. 
In the performed WPI experiments, electronic coherences were induced in a diatomic molecule allowing to study the coupled electronic-vibrational dynamics. 
This is in contrast to pure rovibrational WPI which is not sensitive to dephasing of electronic coherences but in return demands significantly less phase control. 
While other groups have performed electronic WPI on effusive molecular beams\cite{blanchet_temporal_1997, ohmori_real-time_2006}, our work demonstrates for the first time coherent electronic fs spectroscopy of a cold supersonic beam with sufficient resolution to disentangle congested electronic and vibrational states. 
 
The combination of the PM approach with mass-resolved photoion detection allows us to specifically probe Rb*He exciplexes formed at the He droplet surface upon laser excitation. 
We achieve a significant enhancement of the signal-to-noise ratio and an excellent resolution when compared to other  experiments on the same system.   
The WPI data reveals a fully resolved vibrational spectrum for transitions between the $5\Pi_{3/2}$ and the $2^2\Pi$, $4^2\Delta$, $6^2\Sigma$ electronic states. 
A comparison with a spectrum from the diatomic potentials of Pascale clearly demonstrates that the experimental data are well suited to test fine details of pair interaction potentials. So far for the Rb*He molecule there is no set of potentials available to reproduce the details provided by the experiment and we expect our results to stimulate the development of more accurate potentials. 

We note that our experiments provide a new and very promising perspective for applying coherent multidimensional spectroscopy to dilute supersonic beams. 
Particularly the unique capabilities of synthesizing tailor-made complexes in He droplets combined with two-dimensional spectroscopy could provide new insight in fundamental photophysical and photochemical processes. 

\subsection*{Acknowledgements}
L$.$ B$.$ thanks the Evangelisches Studienwerk e$.$V$.$ Villigst for financial support. Furthermore, support by the Deutsche Forschungsgemeinschaft within the IRTG 2079 ''Cold Controlled Ensembles in Physics and Chemistry`` is gratefully acknowledged.

\providecommand*{\mcitethebibliography}{\thebibliography}
\csname @ifundefined\endcsname{endmcitethebibliography}
{\let\endmcitethebibliography\endthebibliography}{}

\end{document}